\documentclass[12pt]{article}
\usepackage{bbm}
\usepackage{epsfig}
\usepackage{array}
\usepackage{float}
\usepackage{dsfont}
\usepackage{amstext}

\usepackage{a4}

\usepackage{a4wide}

\def\ra{\rightarrow}
\def\be{\begin{equation}}
\def\ee{\end{equation}}
\def\gs{\mathrel{
   \rlap{\raise 0.511ex \hbox{$>$}}{\lower 0.511ex \hbox{$\sim$}}}}
\def\ls{\mathrel{
   \rlap{\raise 0.511ex \hbox{$<$}}{\lower 0.511ex \hbox{$\sim$}}}}

\newcommand{\ba}{\begin{array}{c}}
\newcommand{\baz}{\begin{array}{cc}}
\newcommand{\bad}{\begin{array}{ccc}}
\newcommand{\bav}{\begin{array}{cccc}}
\newcommand{\bea}{\begin{equation} \begin{array}{c}}
\newcommand{\eea}{ \end{array} \end{equation}}
\newcommand{\ea}{\end{array}}
\newcommand{\D}{\displaystyle}
\newcommand{\dms}{\mbox{$\Delta m^2_{\odot}$}}
\newcommand{\dma}{\mbox{$\Delta m^2_{\rm A}$}}

\begin{document}

\title{\vspace{.61cm}
\vskip -0.3cm
\hfill {\small arXiv: 0903.4590 [hep-ph]}
\vskip 1.4cm
\Large \bf
Non-Unitary Lepton Mixing Matrix, Leptogenesis and 
Low Energy CP Violation}
\author{Werner Rodejohann$^a$\thanks{email: \tt
werner.rodejohann@mpi-hd.mpg.de} \\ \\ 
{\normalsize \it$^c$Max--Planck--Institut f\"ur Kernphysik,}\\
{\normalsize \it  Postfach 103980, D--69029 Heidelberg, Germany} }
\date{}
\maketitle
\thispagestyle{empty}
\vspace{-0.8cm}
\begin{abstract}
\noindent
It is well-known that for unflavored leptogenesis there is 
in general no connection between low and 
high energy CP violation. We stress that for a non-unitary 
lepton mixing matrix this may not be the case. 
We give an illustrative example for this connection and show 
that the non-standard CP phases that are induced by 
non-unitarity can be responsible for observable effects
in neutrino oscillation experiments as well as for the generation 
of the baryon asymmetry of the Universe. 
Lepton Flavor Violation in decays such as $\tau \ra \mu \gamma$ can
also be induced at an observable level. We also comment on neutrino 
mass limits from leptogenesis, which get barely 
modified in case of a non-unitary mixing matrix. 
\end{abstract}

\newpage

\section{Introduction}

Hands-on beyond the Standard Model physics became 
reality when observations of 
neutrino oscillations showed that neutrino masses are non-zero. 
The most appealing scenario to explain the smallness of 
neutrino masses is the see-saw mechanism \cite{I}, which 
suppresses their mass scale by the presence of new heavy Majorana 
neutrinos. As a bonus, 
there is the possibility that the baryon asymmetry of the Universe is
generated by the out-of-equilibrium decay of these heavy neutrinos, 
the leptogenesis mechanism \cite{lepto}. Unfortunately, 
there is a 
``no connection'' theorem \cite{nolow}, which states 
that low energy CP 
violation in neutrino oscillation experiments 
is independent from the high energy CP violation responsible for 
unflavored leptogenesis. 

In the meanwhile, neutrino physics has entered the 
precision era. In the next decade there will be a plethora of new neutrino oscillation 
and mass-related experiments, whose purpose it is to probe the 
unknown parameters of the neutrino mass matrix and to determine 
the already known ones with high precision \cite{rev}. 
Soon the ``standard picture'' of three active neutrinos whose
mixing is described by a unitary matrix can be put to the test. 
Here we will assume that the standard picture is incomplete, 
in the sense that the lepton mixing matrix deviates from 
being unitary. 
This means that the three active neutrinos $\nu_e$, $\nu_\mu$ and 
$\nu_\tau$ are connected to the mass states $\nu_{1}$, $\nu_{2}$ 
and $\nu_{3}$ via $\nu_\alpha = N_{\alpha i} \, \nu_i$, where 
$N N^\dagger \neq \mathbbm{1}$. As usual, $\alpha = e, \mu, \tau$ 
is the flavor index and $i = 1,2,3$ the mass index. 
This possible feature has recently been discussed by several  
authors 
\cite{NU,SA,CP_Sp,GO,others,Xing_neu}. 
It turns out that many theories 
beyond the Standard Model, which also incorporate massive neutrinos, 
have the capacity to induce a non-unitary PMNS matrix. 
There are various straightforward examples with such a net effect, 
for instance mixing with sterile neutrinos or 
supersymmetric particles, or non-standard 
interactions \cite{NSI,NSI1}. 
Here we will discuss some peculiar implications  
of a non-unitary PMNS matrix. For instance, we note that 
the ``no connection'' theorem between low and high 
energy CP violation no longer holds. We furthermore show that 
the phases associated with the non-unitarity of the 
lepton mixing matrix can be sufficient to generate the 
correct amount of the baryon asymmetry and at the same time 
can lead to spectacular effects in
neutrino oscillation experiments. As one  
example in which this interesting situation may be realized, 
we utilize a simple extension of the see-saw mechanism. 
As a further simple application, we show that lepton flavor 
violating processes such as $\tau \ra \mu \gamma$ can be 
generated at observable levels. Moreover, the non-unitarity of 
the PMNS matrix 
is here shown to lead to maximal values of the decay asymmetry 
which are larger than for a unitary mixing matrix. The small 
impact on neutrino mass limits is discussed.\\ 

The paper is build up as follows: in Section 2 we introduce 
non-unitarity of the PMNS matrix and discuss how it influences 
the see-saw reconstruction. Section 3 deals with non-unitarity  
phenomenology in neutrino oscillations, lepton flavor 
violation and leptogenesis. After investigating neutrino mass 
constraints from leptogenesis in Section 4 we conclude in 
Section 5.

\section{Non-Unitarity, the See-Saw Mechanism and Leptogenesis}
In the conventional see-saw framework there are 
Dirac and Majorana 
mass matrices $m_D$ and $M_R$ in the Lagrangian \cite{I}
\be
{\cal L} = \frac 12 \, \overline{N_R} \, M_R \, N_R^c + 
\overline{N_R} \, m_D \, \nu_L + h.c.
\ee 
Without loss of generality we will assume here that 
$M_R$ is real and diagonal. 
The low energy mass matrix is 
\be
m_\nu = -m_D^T \, M_R^{-1} \, m_D \, . 
\ee 
In the charged lepton basis and the standard picture $m_\nu$ is 
diagonalized by a unitary matrix as 
$U^\ast \, P^\ast \, m_\nu^{\rm diag} \,  P^\dagger \, 
U^\dagger = m_\nu$. The 
Pontecorvo-Maki-Nakagawa-Sakata (PMNS) matrix is $U \, P$, 
where $U$ 
has the standard form 
{\small 
\be \label{eq:Upara} 
U = 
\left( 
\bad  
c_{12}  c_{13} & s_{12}  c_{13} & s_{13}
e^{- i \delta}
\\[0.2cm] 
-s_{12}  c_{23} - c_{12}  s_{23} 
 s_{13}   e^{i \delta} 
& c_{12}   c_{23} - s_{12}   s_{23}  
 s_{13}  e^{i \delta} 
& s_{23}   c_{13} 
\\[0.2cm] 
s_{12}   s_{23} - c_{12}   c_{23}  
 s_{13}  
e^{i \delta} & 
- c_{12}  s_{23} - s_{12}   c_{23}  
 s_{13}   e^{i \delta} 
& c_{23}  c_{13} 
\ea 
\right) .
\ee }
\noindent Here 
$c_{ij} = \cos \theta_{ij}$, $s_{ij} = \sin \theta_{ij}$ 
and the Majorana phases are contained 
in $P = {\rm diag}(1, \, e^{i\alpha} , \, e^{i\beta})$. 

Leptogenesis is a possible consequence of the see-saw mechanism
and generates the baryon asymmetry of the Universe via decays 
of a (usually the lightest) heavy Majorana neutrino into lepton and 
Higgs doublets in the early Universe \cite{D}. 
One distinguishes flavored and unflavored leptogenesis. For  
unflavored leptogenesis, valid for $M_1 \gs 10^{11}$ GeV, 
the flavor of the final state leptons plays no role. Leptogenesis for 
lower values of $M_1$ can be shown to depend on the flavor of 
the final state leptons, and is called flavored leptogenesis 
\cite{flavor_flav}. Here we focus on 
unflavored leptogenesis in which case the decay 
asymmetry is given by 
$$ 
\varepsilon_1 = 
\frac{\D 1}{\D 8 \pi \, v^2} 
\, \frac{\D 1}{\D (m_D m_D^\dagger)_{11}} 
\sum \limits_{j = 2,3}  
 {\rm Im} \left\{ (m_D m_D^\dagger)_{1j}^2 
\right\} \, f(M_j^2 /M_1^2) \, ,
$$
where 
$f(x) \simeq - \frac{3}{2 \sqrt{x}}$ for  
$x \gg 1$, i.e., hierarchical heavy neutrinos. 
The baryon asymmetry of the Universe is 
proportional to the decay asymmetry $\varepsilon_1$. 
It is well-known that in the general case the CP violation 
responsible for unflavored leptogenesis bears no connection 
to low energy lepton mixing angles and CP phases 
(for analyses of low and high energy CP violation 
in case of flavored leptogenesis, 
see e.g.~\cite{sissa0,sissa1}).  
One simple proof of this fact, which we repeat 
here, uses the Casas-Ibarra 
parametrization of the Dirac mass matrix 
in terms of measurable parameters and 
a complex and orthogonal matrix $R$ 
\cite{CI}: 
\be \label{eq:CI}
m_D = i \, \sqrt{M_R} \,  R \, \sqrt{m_\nu^{\rm diag}} 
\,  U^\dagger \, .
\ee
The matrix $R$ contains six free parameters and can be parameterized 
(up to reflections) as
$$
{\small 
R = 
\left( 
\bad  
c_{12}^R  c_{13}^R & s_{12}^R  c_{13}^R & s_{13}^R
\\[0.2cm] 
-s_{12}^R  c_{23}^R - c_{12}^R  s_{23}^R 
 s_{13}^R 
& c_{12}^R  c_{23}^R - s_{12}^R   s_{23}^R  
 s_{13}^R 
& s_{23}^R   c_{13}^R
\\[0.2cm] 
s_{12}^R  s_{23}^R - c_{12}^R   c_{23}^R  
 s_{13}^R & 
- c_{12}^R  s_{23}^R - s_{12}^R   c_{23}^R  
 s_{13}^R 
& c_{23}^R  c_{13}^R
\ea 
\right) ,
}
$$
with complex angles $\theta_{ij}^R$ and 
$c_{ij}^R = \cos \theta_{ij}^R$, $s_{ij}^R = \sin \theta_{ij}^R$. 
The relevant quantity for leptogenesis is then 
$$ 
\ba  
\bav 
m_D m_D^\dagger = &  
\sqrt{M_R} \, R \, \sqrt{m_\nu^{\rm diag}} & \!\! 
 \underbrace{U^\dagger \, U} 
&\!\!  \sqrt{m_\nu^{\rm diag}} \, 
R^\dagger \, \sqrt{M_R} \\ 
& & \!\!\!\! =\mathbbm{1} &  
\ea \\ \hspace{-.8643cm}
= \sqrt{M_R} \, R \, m_\nu^{\rm diag} \, R^\dagger \, \sqrt{M_R} 
\ea
$$ 
Hence, owing to the {\it assumed unitarity} of the PMNS matrix, the low
energy mixing matrix elements (and in particular the CP phases) 
drop out of the expression for the decay asymmetry \cite{nolow}.  
Note further that if $R$ is real then there is no leptogenesis at
all. 

As given in the above expression Eq.~(\ref{eq:Upara}), 
$U$ is indeed manifestly unitary and the ``no connection'' 
theorem is valid. Here we shall assume however that 
the lepton mixing matrix, from now on called 
$N$, is non-unitary. It proves convenient to 
write in the relation $\nu_\alpha = N_{\alpha i} \, \nu_i$, 
which connects flavor and mass states, the non-unitary matrix $N$ 
as \cite{CP_Sp}
\be \label{eq:Ueta}
N = (1 + \eta) \, U_0\,,
\ee   
where $\eta$ is hermitian (containing 6 real moduli 
and 3 phases) and $U_0$ is unitary (containing 3 real moduli 
and 3 phases). Several observables 
lead to 90\% C.L.~bounds on $\eta$ \cite{SA}:  
\be \label{eq:etas}
\left(\bad 
|\eta_{ee}| & |\eta_{e\mu}| & |\eta_{e\tau}| \\
\cdot & |\eta_{\mu\mu}| & |\eta_{\mu\tau}|\\
\cdot &  \cdot & |\eta_{\tau\tau}| 
\ea \right) < 
\left(\bad 
 5.5 \times 10^{-3} & 3.5 \times 10^{-5} & 8.0 \times 10^{-3} \\
\cdot & 5.0 \times 10^{-3} & 5.1 \times 10^{-3} \\
\cdot & \cdot & 5.1 \times 10^{-3}
\ea \right) .
\ee
More importantly for our matters, the possible CP phases of the
elements of $\eta$ ($\eta_{\alpha \beta} = |\eta_{\alpha \beta}| 
\, e^{i \phi_{\alpha \beta}}$ for $\alpha \neq \beta$) 
are not constrained. If $m_\nu$, which is diagonalized by a 
non-unitary mixing matrix, stems from 
the see-saw mechanism, we have now instead of Eq.~(\ref{eq:CI})
\be \label{eq:CI2}
m_D =  i \, \sqrt{M_R} \,  R \, \sqrt{m_\nu^{\rm diag}} 
\,  N^\dagger \, ,
\ee
and thus for leptogenesis 
\be \label{eq:main}
\bav 
m_D m_D^\dagger = &  
\sqrt{M_R} \, R \, \sqrt{m_\nu^{\rm diag}} & \!\!\!\! \!\!\!\! 
\!\!\!\! \!\!\!\! \underbrace{N^\dagger N} 
&\!\!\!\! \!\!\!\! \!\!\!\! \!\!\!\! \sqrt{m_\nu^{\rm diag}} \, 
R^\dagger \, \sqrt{M_R} 
\\
& & \!\!\!\! \!\!\!\!\!\!\!\! 
\simeq 
\mathbbm{1} + 2 \, U_0^\dagger \, \eta \, U_0 \neq \mathbbm{1} &  
\ea 
\ee
We see that leptogenesis is no longer independent on the low 
energy phases. It is sensitive to the phases in $U_0$ as well 
as to the phases in $\eta$. Moreover, in the expression 
Im$\{(m_D  m_D^\dagger)_{1j}^2\}$, which appears in the 
decay asymmetry $\varepsilon_i$, the non-unitarity parameter 
appears in first order! 
In order to underscore the correlation 
between low and high energy CP violation we will 
consider the case of real $R$ from now on. For complex $R$ the 
effect of the non-standard phases can be 
expected to be suppressed by the smallness of the 
$\eta_{\alpha\beta}$. 
Nevertheless, as we will show below, the non-standard 
CP phases included in $\eta$ are sufficient to 
generate the baryon asymmetry of
the Universe via the leptogenesis mechanism\footnote{Let us note 
that flavored leptogenesis, 
where in general 
contributions from low energy phases can 
be expected \cite{sissa0,sissa1}, will of course receive 
contributions from the non-standard phases as well.}. 
In addition, 
spectacular effects in neutrino oscillation experiments 
can be induced \cite{CP_Sp,GO}.\\  
 
First of all we will outline a possible see-saw extension which 
incorporates the situation we are after. 
The intrinsic non-unitarity in general 
see-saw scenarios has first been noted in \cite{Jose}. 
Let us stress here that also the conventional see-saw mechanism
implies that the PMNS matrix is non-unitarity. However, 
for the natural case of heavy Majorana neutrino masses 
(which we require in order to have thermal leptogenesis) 
this effect 
is way too tiny to lead to any observable signature, 
see e.g.~\cite{Xing_neu}. 
Note that in order to 
link the lepton mixing matrix to leptogenesis via the Casas-Ibarra 
parametrization, there should be no sizable contribution 
to $m_\nu$ and leptogenesis other than the usual 
see-saw terms. Hence we need to decouple 
the source of unitary violation from these terms, but still allow 
some mixing of the light neutrinos with new physics, 
thereby creating non-unitarity 
in the low energy mixing matrix. 
Consider the see-saw 
mechanism extended by an additional singlet sector, 
leading to a $9\times9$ mass matrix 
$$ 
{\cal L} = \frac 12 \, 
\left( \overline{\nu_L^c} \, ,~\overline{N_R} \, 
,~\overline{X}\right)
\left( 
\bad
0 & m_D^T & m^T \\
m_D & M_R & 0 \\
m & 0 & M_S
\ea
\right) 
\left( 
\ba
\nu_L \\ N_R^c \\ X^c
\ea
\right) + h.c., 
$$
where the upper left block is the usual see-saw. 
We can diagonalize it with a unitary matrix 
${\cal U}$ defined as 
$$ 
\ba 
{\cal U} = 
\left( 
\bad
\tilde N & S & A \\ 
T & V & D \\
B & E & W 
\ea
\right) ,\mbox{ where } \\ 
{\cal U}^T 
\left( 
\bad
0 & m_D^T & m^T \\
m_D & M_R & 0 \\
m & 0 & M_S
\ea
\right) 
{\cal U}
= 
\left( 
\bad
m_\nu^{\rm diag} & 0 & 0 \\ 
0 & M_R & 0 \\ 
0 & 0 & M_S^{\rm diag}
\ea \right) .
\ea
$$
The usual see-saw terms take their usual magnitudes, 
$M_R \sim 10^{15}$ GeV and $m_D \sim 10^2$ GeV. 
Consequently, $S$ and $T$ are of order $m_D/M_R \sim 10^{-13}$. 
Assume now that the rotation that eliminates the 
13-entry $m$ (corresponding to the matrices $A$ and $B$) 
is of order $10^{-2}$. This rotation introduces
additional terms to the low energy neutrino mass matrix 
of order $m \times 10^{-2}$ and $M_S \times 10^{-4}$. 
We need to suppress these entries with respect to $m_D^2 /M_R$. 
For instance, if $m$ is of order $10^{-13}$ GeV and 
$M_S$ of order $10^{-15}$ GeV, 
it is easy to see that $A$ and $B$ are of 
order $10^{-2}$. 
These features together with the mentioned magnitudes of $S$ and $T$ 
are enough to obtain from the upper left and upper middle elements 
of the mass matrix:  
\be \label{eq:s}
m_\nu^{\rm diag} = - \tilde N^T \, m_D^T \, M_R^{-1} \, 
m_D \, \tilde N \, .
\ee
This shows that $N = (\tilde{N}^\dagger)^{-1}$ 
is the leptonic mixing 
matrix, since there is no other sizable contribution to 
the mass term of the light 
active neutrinos.  
$N$ is non-unitary, because the 11-element of 
${\cal U} {\cal U}^\dagger$ gives the constraint 
$(N^\dagger)^{-1} N^{-1} + S S^\dagger + A A^\dagger  
= \mathbbm{1}$. 
Because the non-unitary contribution from $S$ 
is extremely tiny, we can neglect it. 
Note that in this case Eq.~(\ref{eq:Ueta}) and 
$(N^\dagger)^{-1} N^{-1} + S S^\dagger + A A^\dagger  
= \mathbbm{1}$ imply 
$2\eta \simeq  A A^\dagger$.  
The unitarity 
violation is therefore governed by the magnitude 
of the elements of the matrix $A$, which is of 
order $10^{-2}$, thereby possibly 
reaching the limits on $\eta$ given above. 

Within this situation we have sketched how to keep 
the usual see-saw mechanism with unmodified leptogenesis 
(note that $M_R$ does not couple to the new singlets)\footnote{There is a suppressed contribution to the 
decay asymmetry in which two mass insertions mediated by the 
$\nu_L$-$X$-vertex show up in the usual self-energy contribution.}
but to induce a sizably non-unitary lepton mixing matrix $N$. 
Note that the hierarchies in the individual mass matrices should 
not be too extreme in order for this mechanism to work. 
Nevertheless, we are satisfied that there are indeed frameworks 
which allow us to apply Eq.~(\ref{eq:main}). 
Another example could be non-standard interactions 
mediated by higher dimensional operators which lead to
non-unitarity \cite{NSI1}. The new 
heavy particles introducing these operators 
are supposed to not influence the see-saw formula 
for $m_\nu$ and leptogenesis, 
which is a constraint on
those scenarios. We will however 
continue our analysis independent of any possible underlying 
setup and simply analyze the consequences of Eq.~(\ref{eq:main}).\\

\section{Low Energy Phenomenology of Non-Unitarity and Leptogenesis}
One interesting aspect of non-unitarity of the PMNS matrix lies 
in Lepton Flavor Violation (LFV). 
It is well-known that even for the unitary case 
massive neutrinos can induce LFV (although at unobservably small
levels) in decays such as 
$\alpha \ra \beta \, \gamma$ with $(\alpha, \beta) = 
(\tau, \mu)$, $(\tau, e)$ or  $(\mu, e)$. 
The branching ratio normalized to 
the leading decay in charged leptons and neutrinos 
is given by \cite{lfv0}
\be \label{eq:BR}
\frac{{\rm BR}(\alpha \ra \beta  \gamma)}
{{\rm BR}(\alpha \ra \beta  \, 
\overline{\nu}  \nu)} 
= \frac{3 \, \alpha}{32 \, \pi} 
\left| 
U_{\alpha i} \, U_{i \beta}^\dagger \, f(x_i) 
\right|^2 \, , 
\ee
where the light neutrino $m_i$ masses appear in $x_i = m_i^2 /m_W^2$  
($m_W$ being the mass of the $W$) and the loop function is 
$$
f(x) = \frac 13 \, \frac{ 
10 - 43 \, x + 78 \, x^2 - 49 \, x^3 + 4 \, x^4 + 18 \, 
x^3 \, \ln x}{(1 - x)^4} \, .
$$ 
Because $x \ll 1$ one has $f(x) \simeq 10/3 - x $ and the 
first order term in Eq.~(\ref{eq:BR}) is absent when summed 
over $i$. This GIM suppression 
mechanism leading to unobservably small branching ratios 
proportional to $m_i^4/m_W^4$ is a consequence of the 
unitarity of the PMNS matrix. 

Now consider unitarity violation: the formula for the 
branching ratio $\alpha \ra \beta \, \gamma$ is
the same as above in Eq.~(\ref{eq:BR}), except that now $U$ is to be
replaced with $N$. Since $N$ is not unitary the first order 
term from $f(x)$ may be the leading contribution and the branching
ratio is 
\be \label{eq:BRNU}
\frac{{\rm BR}(\alpha \ra \beta  \gamma)}
{{\rm BR}(\alpha \ra \beta  \, 
\overline{\nu}  \nu)} 
\simeq \frac{100 \, \alpha}{96 \, \pi} 
\left| 
(N N^\dagger)_{\alpha \beta}
\right|^2 \, . 
\ee
The fact that the experimental limit  
BR$(\mu \ra e \gamma) < 1.2 \cdot 10^{-11}$ \cite{meg}, 
is almost four orders of magnitude stronger than the limits 
BR$(\tau \ra e \gamma) < 3.3 \cdot 10^{-8}$ or 
BR$(\tau \ra \mu \gamma) < 4.4 \cdot 10^{-8}$ \cite{BR}
is the reason why the limit on $|\eta_{e \mu}|$ is almost two 
orders of magnitude better than the ones on 
$|\eta_{e \tau}|$ or $|\eta_{\mu\tau}|$. 
We consider from now on only the term 
$|\eta_{\mu\tau}| \, e^{i \phi_{\mu\tau}}$ to 
illustrate our points with a simple example.
The simple result is 
\be \label{eq:BRsimple}
{\rm BR}(\tau \ra \mu \gamma) \simeq 
{\rm BR}(\tau \ra \mu  \, \overline{\nu}  \nu)  \, 
\frac{25 \, \alpha}{6 \, \pi} \, 
\, |\eta_{\mu \tau}|^2 \, .
\ee
Note that with $N = (1 + \eta) \, U_0$ the parameters in $U_0$ 
do not appear in $N N^\dagger$. 
Using ${\rm BR}(\tau \ra \mu  \, \overline{\nu}  \nu) 
= 0.1736$ and ${\rm BR}(\tau \ra \mu  \gamma) 
< 4.4 \cdot 10^{-8} $ \cite{BR} one can obtain  
the constraint $|\eta_{\mu \tau}| < 5.1 \cdot 10^{-3}$ 
quoted above.  Alternatively, if $|\eta|$ is 
given, or constrained to lie in a certain range, one can 
predict the rate of $\tau \ra \mu  \gamma$. Indeed, rewriting the last
expression gives a useful estimate:  
\be \label{eq:BRes}
{\rm BR}(\tau \ra \mu \gamma) 
\simeq 
4.2 \cdot 10^{-10} \,  
\left(\frac{\D 
|\eta_{\mu \tau}|}{\D 5 \cdot 10^{-4}}\right)^2 \, .
\ee
It has been estimated that improvements on the experimental 
limits of ${\rm BR}(\tau \ra \beta \gamma)$ by 
one to two orders of magnitude are possible \cite{es}. 
Sensitivities around $2 \cdot 10^{-9}$ correspond to  
values of $|\eta_{\mu \tau}|$ around $1.1 \cdot 10^{-3}$.\\ 

As a next application of unitarity violation we consider 
effects in neutrino oscillation experiments: 
it is known \cite{CP_Sp,GO} that non-unitarity of 
the PMNS matrix can lead to CP asymmetries in neutrino 
oscillations, which are dramatically different from the ones in 
the standard case. 
One defines asymmetries 
\be 
A_{\alpha \beta}^{\rm SM} = \frac{
P_{\alpha \beta}^{\rm SM} - 
P_{\overline{\alpha} \overline{\beta}}^{\rm SM}}
{P_{\alpha \beta}^{\rm SM} + 
P_{\overline{\alpha} \overline{\beta}}^{\rm SM}}
\ , 
\ee
where $P_{\alpha \beta}^{\rm SM}$ is the standard 
(vacuum) oscillation probability for the channel 
$\nu_\alpha \rightarrow \nu_\beta$ and 
$P_{\overline{\alpha} \overline{\beta}}^{\rm SM}$ 
the corresponding probability for 
anti-neutrinos:  
\be \label{eq:PSM}
P_{\alpha \beta}^{\rm SM} = \left| 
\sum_j U_{\beta j} \, U_{\alpha j}^\ast \, e^{-i m_j^2 L/(2E)} \right|^2 
\ee
and $P_{\overline{\alpha} \overline{\beta}}^{\rm SM} 
= P_{\alpha \beta}^{\rm SM}(U \rightarrow U^\ast)$. 
Survival probability asymmetries $A_{\alpha \alpha}^{\rm SM}$ 
are zero and $A_{\alpha \beta}^{\rm SM} = -A_{\beta \alpha}^{\rm
SM}$. 
We focus from now on on one
particular setup, in which we can simplify the expressions. 
We assume negligible matter effects and $\Delta_{21} \ll
|\Delta_{31}|$, where $\Delta_{ij} = \Delta m^2_{ij} \, L/(2 E)$ with 
$\Delta m^2_{ij} = m_i^2 - m_j^2$. 
This may be realized in a high-energy, short-baseline 
neutrino factory, which has been discussed in the framework of 
non-unitarity effects in oscillations first in Ref.~\cite{CP_Sp}. 
Following those authors we will choose as an example 
a muon energy $E_\mu = 50$ GeV (hence $\langle E_{\nu_\mu} \rangle 
= \frac{7}{10} \, E_\mu)$ and baseline $L = 130$ km. 
It was found that the $\mu$--$\tau$ channel is 
particularly powerful to 
detect new physics. In the unitary case and with the simplifications
given above one finds for $\sin^2 \theta_{12} = \frac 13$ and 
$\sin^2 \theta_{23} = \frac 12$ that 
\be
A_{\mu \tau}^{\rm SM} \simeq \frac{2\sqrt{2}}{3} \, \Delta_{21} 
\, |U_{e3}| \, \sin \delta \, .
\ee
Numerically, for $\theta_{13} = 0.1$ one finds from the full 
expression that $|A_{\mu \tau}^{\rm SM}| \ls 7 \times 10^{-5}$.

For a non-unitary lepton mixing matrix the oscillation 
probabilities are   
$$ 
P_{\alpha \beta}^{\rm NU} = 
\frac{
\left|\sum_j N_{\beta j} \, N_{\alpha j}^\ast \, e^{-i m_j^2 L/(2E)}
\right|^2}
{
(N N^\dagger)_{\alpha\alpha}  \, 
(N N^\dagger)_{\beta\beta}  
} 
\equiv 
\frac{\hat{P}_{\alpha \beta}^{\rm NU}
} {
(N N^\dagger)_{\alpha\alpha}  \, 
(N N^\dagger)_{\beta\beta}  
} \, 
.
$$
For anti-neutrinos one has 
$P_{\overline{\alpha} \overline{\beta}}^{\rm NU} 
= P_{\alpha \beta}^{\rm NU}(N \rightarrow N^\ast)$. 
As can be seen, 
there is just as for ``normal'' oscillations no sensitivity on the 
``Majorana phases'' which appear on the right of $U_0$. 
The CP asymmetries are  
\be \label{eq:ANU}
A_{\alpha \beta}^{\rm NU} = 
\frac{
\hat{P}_{\alpha \beta}^{\rm NU} - 
\hat{P}_{\overline{\alpha} \overline{\beta}}^{\rm NU}
}
{\hat{P}_{\alpha \beta}^{\rm NU} + 
\hat{P}_{\overline{\alpha} \overline{\beta}}^{\rm NU}} 
\, .
\ee
The different $A_{\alpha \beta}^{\rm NU}$ are 
in general independent of each other, though 
(just as in the standard case) 
survival probability asymmetries 
$A_{\alpha \alpha}^{\rm NU}$ are zero and 
$A_{\alpha \beta}^{\rm NU} = -A_{\beta \alpha}^{\rm NU}$.  
Note that the asymmetries were constructed in terms of 
``bare'' probabilities, i.e., probabilities stripped of 
their normalization factors $(N N^\dagger)_{\alpha\alpha}$.

For simplicity, we will assume from now on that $U_0$, 
which can be parameterized in analogy to Eq.~(\ref{eq:Upara}), 
is characterized by $\sin^2 \theta_{12} = \frac 13$ and 
$\sin^2 \theta_{23} = \frac 12$. This is motivated by the 
facts that at leading order $U_0$ can be identified with the PMNS
matrix, and that this matrix is at leading order well described 
\cite{bari} by tri-bimaximal mixing.  
The CP asymmetry in the 
$\mu$--$\tau$ sector is then 
\be \label{eq:Amt}
A_{\mu\tau}^{\rm NU} \simeq  
\frac{-4 \, |\eta_{\mu\tau}|}{\cos^2 \theta_{13}}
\cot \frac{\Delta_{31}}{2} \, \sin
\phi_{\mu\tau}
\, . 
\ee
Hence, in the experimental setup under consideration 
$|A_{\mu\tau}^{\rm NU}|$ can easily exceed the maximal value of 
$|A_{\mu\tau}^{\rm SM}| \ls 7 \times 10^{-5}$ given above 
\cite{CP_Sp,GO}.\\

Returning to leptogenesis, the baryon asymmetry should lie in the 
interval $(8.75 \pm 0.23) \times 10^{-11}$ and is given by \cite{D}
$ Y_B \simeq 
1.27 \times 10^{-3} \, \varepsilon_1 \, \eta(\tilde{m}_1)$, 
where $g_\ast \simeq 106$, $\tilde{m}_1 = (m_D 
m_D^\dagger)_{11}/M_1$ and 
$ \eta(x) \simeq 
1/( (8.25 \times 10^{-3}~{\rm eV})/x + 
(x/(2 \times 10^{-4}~{\rm eV})^{1.16} 
)$ describes the wash-out. 
The in general complex and orthogonal matrix 
$R$ in the parametrization of $m_D$ from Eq.~(\ref{eq:CI2}) is 
here chosen real (to forbid unflavored leptogenesis in case of a 
unitary PMNS matrix) and described by three real Euler 
angles $\theta_{ij}^R$ with $ij = 12, 13, 23$. 
The light neutrino masses are normally ordered with 
$m_1 = 10^{-3}$ eV, $m_2^2 - m_1^2 = \dms$ and 
$m_3^2 - m_1^2 = \dma$, where 
$\dms = 7.67 \times 10^{-5}$ eV$^2$ and 
$\dma = 2.39 \times 10^{-3}$ eV$^2$ \cite{bari}. 
Neglecting $m_1$ and $\theta_{13}$ we find 
\bea \nonumber 
\varepsilon_1 \simeq \D 
\sqrt{\frac{3}{8 \pi^2 }} 
\, \frac{M_1 \, \sqrt{m_2 \, m_3}}{v^2} 
\frac{(m_2 - m_3) \, c_{13}^R \, s_{12}^R \, s_{13}^R}
{m_2 \, R_{12}^2 + m_3 \, R_{13}^2}  
|\eta_{\mu \tau} | \, \sin \phi_{\mu\tau} \, ,
\eea
where $R_{ij}$ are the elements of $R$ and $s_{ij}^R = \sin
\theta_{ij}^R$, $c_{ij}^R = \cos \theta_{ij}^R$. 
Hence, just as the oscillation CP asymmetry 
in the $\mu$--$\tau$ sector in Eq.~(\ref{eq:Amt}), 
and just as expected from Eq.~(\ref{eq:main}), 
the decay asymmetry depends linearly on 
$|\eta_{\mu \tau} | \, \sin \phi_{\mu\tau}$. 
Note that $\varepsilon_1 = 0$ for $\eta_{\mu \tau} = 0$. 
\begin{figure}[t]
\begin{center}
\includegraphics[width=12cm,height=9cm]{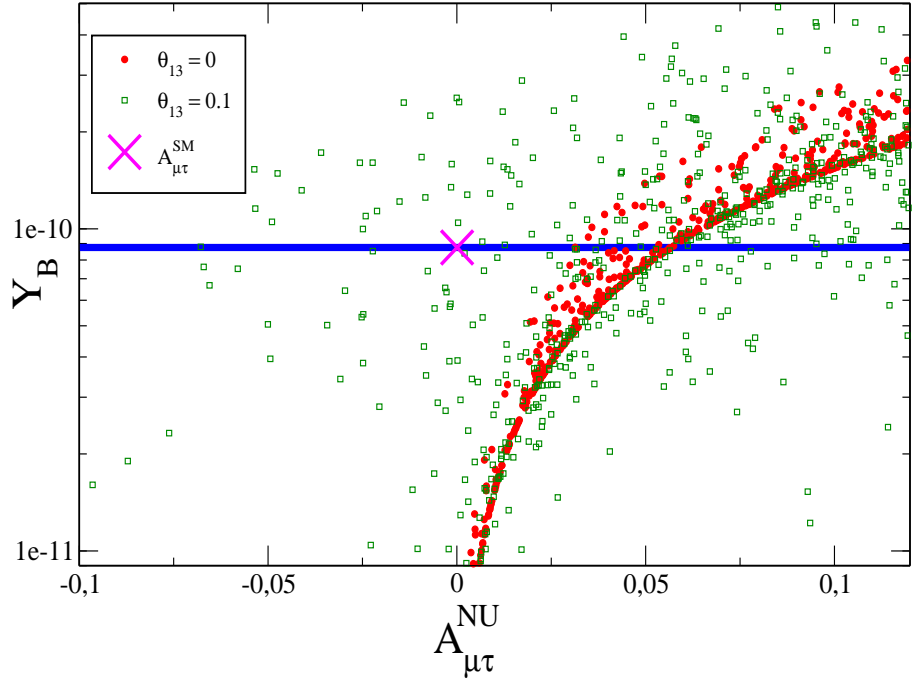}
\end{center}\vspace{-.5cm}
\caption{\label{fig:1}Scatter plot of the CP asymmetry in neutrino 
oscillations for a 50 GeV neutrino factory with baseline 130 km 
against the baryon asymmetry of the Universe. 
We have chosen $\theta_{12}^R = 1.1$, $\theta_{23}^R = 0.3$ and 
$\theta_{13}^R = 0.8$. 
The blue horizontal line is the observed value of $Y_B$. The maximal 
value for $|A_{\mu \tau}|$ in the Standard Model is 
marked by a magenta cross and is 
$7 \times 10^{-5}$ for $\theta_{13} = 0.1$.}
\end{figure}
Hence, 
the dependence of $\varepsilon_1$ on the ``standard phase'' $\delta$ 
should be at most proportional to 
$|\eta_{\mu\tau}| \, \theta_{13} \, \sin \delta$, 
which is indeed the case. 
For $M_1$, $m_2$ and $m_3$ given as above and for, say,  
$\theta_{12}^R = \theta_{13}^R = \pi/4$, the order of magnitude 
of the decay asymmetry is $\varepsilon_1 \simeq -10^{-4} \, 
|\eta_{\mu \tau} | \, \sin \phi_{\mu\tau}$. 
The wash-out parameter $\tilde{m}_1$ is of order 
$10^{-2}$ eV. This means we are in the favorable ``strong
wash-out regime'',  in which the final baryon asymmetry 
has very little to no dependence on the initial conditions 
(i.e., the initial abundance of the heavy neutrinos) \cite{D}.  
In this regime $\eta(\tilde{m}_1)$ is of order $10^{-2}$, and
in total we end up with 
\be
Y_B \sim {\rm few}\, 10^{-9} \, |\eta_{\mu \tau} | 
\, \sin \phi_{\mu\tau} \, ,
\ee
which for $|\eta_{\mu \tau}|$ of order few times $10^{-3}$ is 
in reasonable agreement with the observed range it 
should lie in. A more detailed numerical study using no approximations
results in Fig.~\ref{fig:1}, which illustrates the interplay 
of low and 
high energy CP violation for a (by no means special) point in 
the parameter space spanned by $R$. 
We have varied $|\eta_{\mu\tau}|$ 
and $\phi_{\mu\tau}$ within their allowed ranges 
and took as heavy neutrino parameters 
$M_1 = 2.5 \cdot 10^{12}$ GeV, 
$M_2 = 2 \cdot 10^{13}$ GeV and $M_3 = 10^{14}$ GeV. 
Note that $M_1$ is chosen such that flavor effects in 
leptogenesis play no role.  
We stress again that in this situation a unitary PMNS matrix 
forbids the generation of a baryon asymmetry via leptogenesis. 
The correct value of the baryon
asymmetry can be generated with values of 
$A_{\mu\tau}^{\rm NU}$ 
almost four orders of magnitude above the SM value. 
Note that we have plotted only positive values of $Y_B$. We also 
show in Fig.~\ref{fig:1} the case of $\theta_{13} = 0.1$, in which
case we have also varied $\delta$.  

\begin{figure}[t]
\begin{center}
\includegraphics[width=12cm,height=9cm]{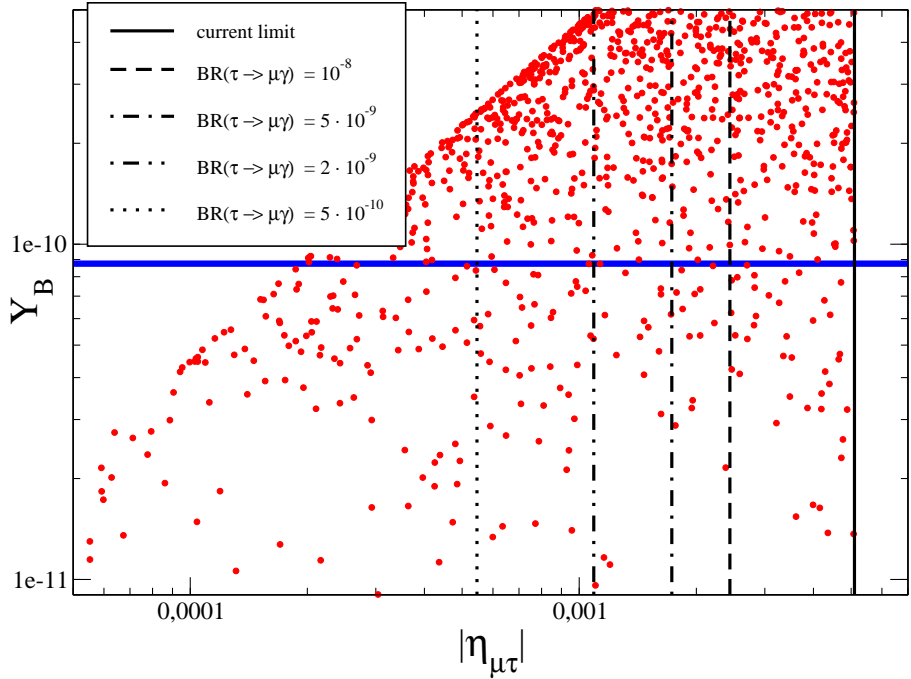}
\end{center}\vspace{-.5cm}
\caption{\label{fig:2}Scatter plot of $|\eta_{\mu\tau}|$
against the baryon asymmetry of the Universe for the same parameters 
as in Fig.~\ref{fig:1} ($\theta_{13} = 0$). Certain values of 
$|\eta_{\mu\tau}|$ which correspond to certain observable benchmark 
values of the branching ratio for the decay 
$\tau\ra \mu \gamma$ 
have also been indicated.}
\end{figure}

It turns out that 
$|\eta_{\mu\tau}| \gs 10^{-4}$ in order to 
generate a sufficient 
$Y_B$. Interestingly, these are values for 
which future facilities will be able to 
probe the non-standard CP phases \cite{CP_Sp}. 
In Fig.~\ref{fig:2} we show $|\eta_{\mu\tau}|$ against 
the baryon asymmetry of the Universe for the case of 
$\theta_{13} = 0$ and everything else as for Fig.~\ref{fig:1}. 
We have also indicated certain values of $|\eta_{\mu\tau}|$ which
correspond to certain observable benchmark values of the branching 
ratio for the decay $\tau\ra \mu \gamma$.   
 Though this is a straightforward and rather simple application,
it serves to illustrate nicely the rich phenomenology of unitarity 
violation.

\section{Mass Limits from Leptogenesis} 
\begin{figure}[t]
\begin{center}
\includegraphics[width=6.96cm,height=5.64cm]{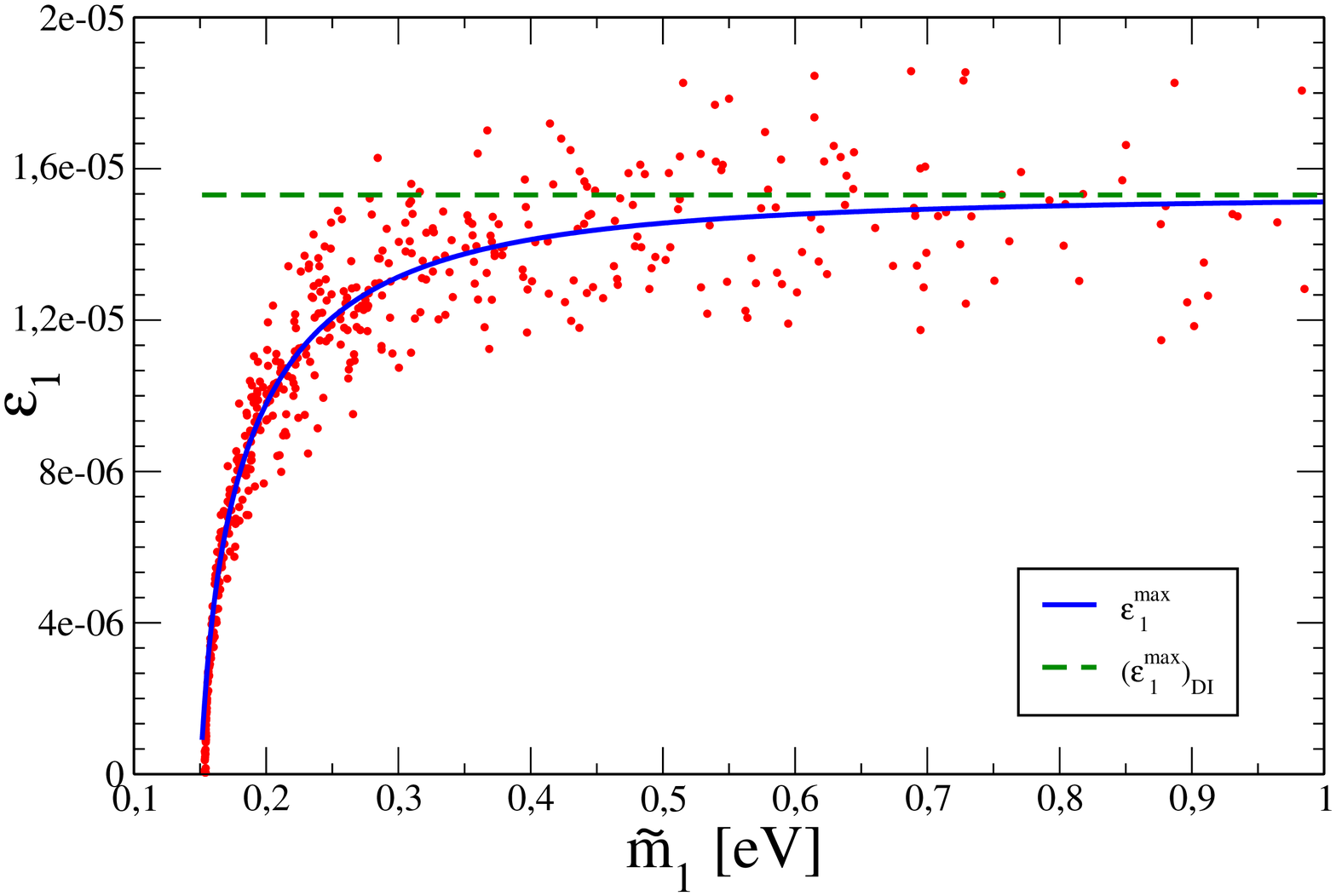}
\includegraphics[width=6.96cm,height=5.64cm]{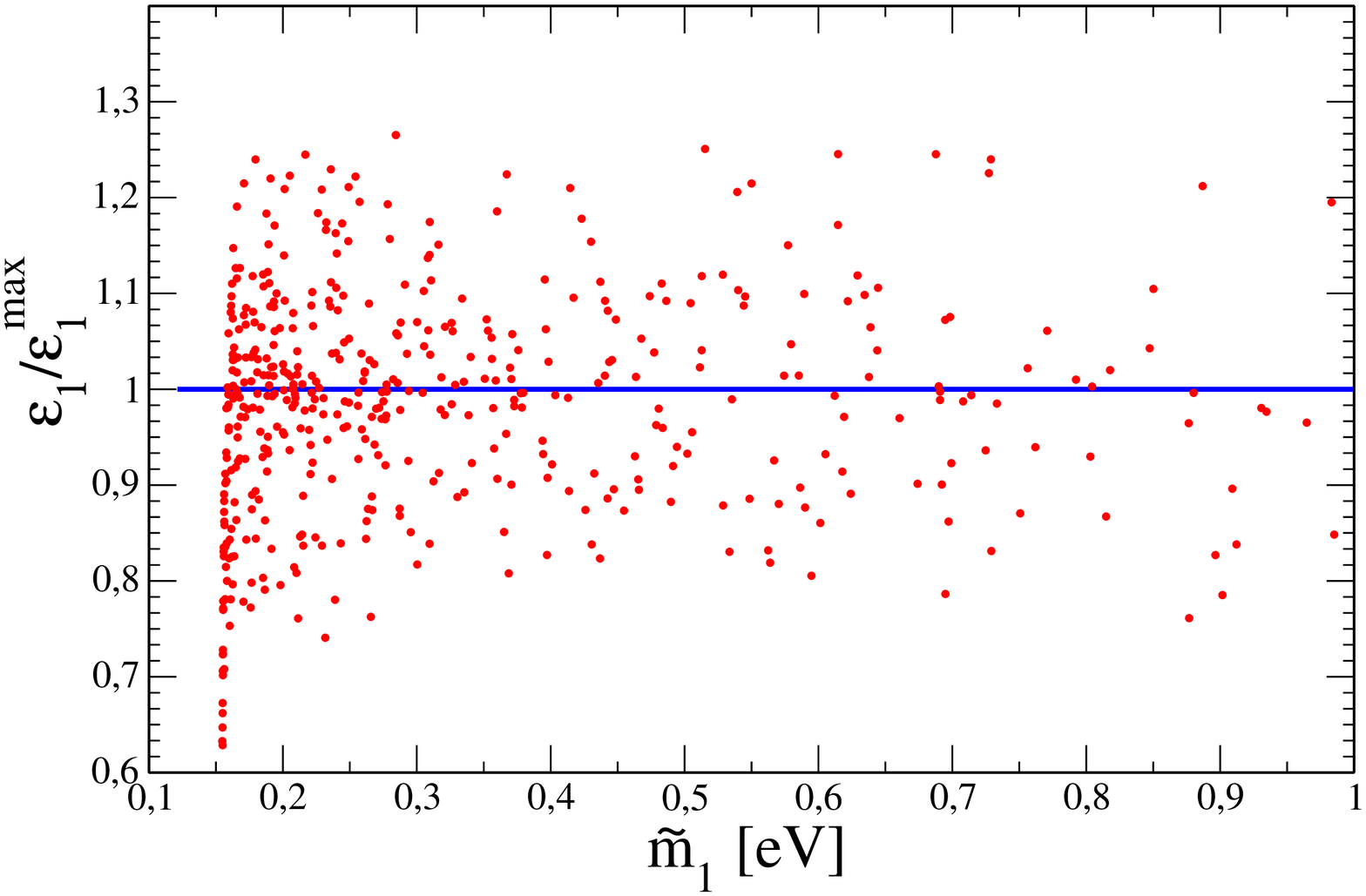}
\end{center}\vspace{-.5cm}
\caption{\label{fig:3}Scatter plot of the decay asymmetry 
$\varepsilon_1$ against $\tilde{m}_1$ (left) and the ratio of 
$\varepsilon_1$ against its maximal value in case of a unitary PMNS
matrix (right).}
\end{figure}

Finally, let us note that the upper limit on the decay asymmetry 
$\varepsilon_1$ is also modified if the PMNS matrix is not unitary. 
A close numerical inspection of the situation reveals 
that for a lightest 
neutrino mass of 0.15 eV, one can exceed the 
upper bound on $\varepsilon_1$ (obtained for a unitary PMNS matrix) 
as a function of $\tilde{m}_1$ by roughly 30 \%. Analytically, 
in the unitary case and with the parametrization of $R$ from above one 
can choose \cite{PdB} $\theta_{12}^R = 0$ and 
$\theta_{13}^R = \rho_{13} + i \, \sigma_{13} = \pi/4 
+ i \, \sigma_{13}$ with real $\sigma_{13}$ to find that 
$$ 
\varepsilon_1 = \frac{3 \, M_1}{16 \, \pi \, v^2} 
\, (m_3 - m_1) \, \tanh 2 \sigma_{13} \, 
$$ 
approaches the Davidson-Ibarra bound \cite{DI}  
$\varepsilon_1^{\rm DI} = \frac{3 \, M_1}{16 \, \pi \, v^2} 
\, (m_3 - m_1)$ in the limit of large $\sigma_{13}$. Ref.\
\cite{hambye} obtained a more general limit as a function of 
$m_1$, $\tilde{m}_1$ and $m_3$: 
\be \label{eq:hambye}
|\varepsilon_1 | \le \frac 12 \, \varepsilon_1^{\rm DI} \, 
\sqrt{1 - ((1 - a) \, \tilde{m}_1/(m_3 - m_1))^2} \, 
\sqrt{(1 + a)^2 - ((m_3 + m_1)/\tilde{m}_1)^2} \, ,
\ee
with 
\be
a = 2 \, {\rm Re} \left(\frac{m_1 \, m_3}{\tilde{m}_1^2} \right)^{1/3}
\left(
-1 - i \, \sqrt{\frac{(m_1^2 + m_3^2 + \tilde{m}_1^2)^3}{27 \, 
m_1^2 \, m_3^2 \, \tilde{m}_1^2}} 
\right)^{1/3} \, .
\ee
Now we consider modifications of these limits in case of 
a non-unitary PMNS matrix. 
As usual, we consider only $|\eta_{\mu\tau}| \, e^{i\phi_{\mu\tau}}$, 
while choosing in $U_0$ the angles to be $\theta_{23}=\pi/4$, 
$\theta_{13} = 0$, $\theta_{12} = \sin^{-1} \sqrt{1/3}$ 
and all phases zero, and finally we find 
\be \label{eq:epsNU}
\ba \D 
\varepsilon_1 \simeq \frac{3 \, M_1}{16 \, \pi \, v^2} 
\, (m_3 - m_1) \, \tanh 2 \sigma_{13} \\ \D 
+ \frac{M_1}{8 \, \pi \, v^2} \, |\eta_{\mu \tau}|   
(m_1 + 3 \, m_3) \, \cos \phi_{\mu\tau} \, 
\tanh 2 \sigma_{13}
 \, .
\ea 
\ee 
We have given here only the leading term proportional to 
$|\eta_{\mu \tau}|$. 
The ratio of this term to the zeroth order term is smaller than 
$$  
\frac{\frac 18 \, (m_1 + 3 \, m_3) \, |\eta_{\mu\tau}|}
{\frac{3}{16} \, (m_3 - m_1)} \simeq 
\frac{16}{3} \, \frac{m_1^2}{\dma} \, |\eta_{\mu\tau}| 
\ls 0.25 \, ,
$$
where we have used that for $m_1 = 0.15$ eV the neutrinos are
quasi-degenerate and $m_3 - m_1 \simeq (m_3^2 - m_1^2)/(2\,m_1)$. 
 This estimate almost fully explains the 30\% effect 
seen in the numerical analysis. 
The reason for the comparably large contribution from the small 
term $|\eta_{\mu\tau}|$ is that the zeroth order expression 
$\varepsilon_1^{\rm DI}$ goes to zero for $m_3 \ra m_1$, 
whereas the terms proportional to $|\eta_{\mu\tau}|$ do not. 
We show in Fig.~\ref{fig:3} scatter plots of the 
decay asymmetry against $\tilde{m}_1$. The solid (blue) 
line in the left plot shows the exact upper bound from 
Eq.~(\ref{eq:hambye}), the dashed (green) line shows the 
Davidson-Ibarra bound. In the right plot of Fig.~\ref{fig:3} 
the ratio of $\varepsilon_1$ to its maximal value
Eq.~(\ref{eq:hambye}) in case of a unitary mixing matrix is shown. 
We have chosen $\theta_{12}^R = 0$, 
$\theta_{13}^R = \rho_{13} + i \, \sigma_{13} = \pi/4 
+ i \, \sigma_{13}$, a normal mass ordering with 
$m_1 = 0.15$ eV, 
$\theta_{23}=\pi/4$, $\theta_{13} = 0$, 
$\theta_{12} = \sin^{-1} \sqrt{1/3}$ and all phases in $U_0$ zero. 
As seen from the plots, the non-unitarity of the mixing 
matrix can violate the upper bound on $\varepsilon_1$ by 30 \%, as
mentioned above.

From a limit on $\varepsilon_1$ it is possible to get limits 
on the light and heavy neutrino masses 
from a detailed analysis of the wash-out \cite{hambye,BBP}. 
For instance, Ref.~\cite{hambye} found that a limit of 
$m_1 \le 0.15$ eV is implied, and Ref.~\cite{BBP} quotes 
$M_1 \ge 4 \cdot 10^8$ GeV. 
We note here that these limits may get modified by the presence 
of non-unitarity, simply because the upper limit on 
$\varepsilon_1$ is modified. 
A detailed analysis is beyond the scope of this
letter, but we can estimate the effect. 
We note first that in contrast to 
the contribution to $\varepsilon_1$, the corrections from  
$|\eta_{\mu\tau}|$ to the wash-out parameter $\tilde{m}_1$ are not 
enhanced.  
The wash-out therefore proceeds to good precision 
in the same manner as for a unitary PMNS matrix and thus the limit 
on $m_1$ is modified with good precision by the same amount 
as the limit on $\varepsilon_1$ is modified. 
 Nevertheless, the upper limit on the light neutrino mass, 
which is roughly proportional to $|\varepsilon_1^{\rm max}|^{0.25}$ 
\cite{priv} is not substantially different than before, 
thus $m_1 \ls 0.16$ eV instead of 0.15 eV. In addition, given 
the uncertainties \cite{hambye} of such bounds, the modification 
due to unitarity violation is not of much significance. 
The lower limit on the lightest heavy neutrino is inversely 
proportional to $|\varepsilon_1^{\rm max}|$, if hierarchical light 
neutrinos are assumed. In this limit, however, 
the impact of non-unitarity on the decay asymmetry is suppressed 
by the smallness of the elements of $\eta$, and hence limits 
on $M_1$ are  basically not modified.

\section{Summary}
The possible non-unitarity of the lepton mixing matrix has several
consequences in phenomenology. We have illustrated here in 
particular that the non-standard CP phases as induced by 
non-unitarity 
have the capacity to lead to observable effects in neutrino 
oscillation experiments as well as 
to a successful generation of the baryon asymmetry of the 
Universe via leptogenesis. This is true even for situations 
in which a unitary lepton mixing matrix would 
lead to no leptogenesis at all. 
The usual ``no connection'' theorem of low and high energy 
CP violation in case of unflavored leptogenesis 
is thus avoided. 
Lepton Flavor Violation in charged lepton decays such as 
$\tau \ra \mu \gamma$ can at the same time be induced at 
an observable level. 
The neutrino mass constraints from leptogenesis are 
only slightly modified. 
We have discussed an extension of the see-saw 
mechanism which can incorporate this framework, but our results 
will apply also for other scenarios as long as 
the source of non-unitarity is decoupled from leptogenesis and from 
$m_\nu = -m_D^T \, M_R^{-1} \, m_D$.\\[0.5cm]

\begin{center}
{\bf Acknowledgments}
\end{center}
We thank Pasquale di Bari, Fedor Bezrukov, Serguey Petcov 
and especially Tom Underwood for helpful discussions. 
This work was supported by the ERC under the Starting Grant 
MANITOP and by the DFG in the Transregio 27.

\end{document}